\documentclass[12pt]{iopart}

\usepackage{iopams}  
\usepackage{natbib} 
\expandafter\let\csname equation*\endcsname\relax
\expandafter\let\csname endequation*\endcsname\relax
\usepackage{amsmath}
\usepackage{amsmath}
\usepackage{graphicx}
\usepackage{booktabs}
\usepackage[final, commandnameprefix=ifneeded]{changes}
\usepackage{subcaption}

\providecommand{\newblock}{\hskip .11em\@plus.33em\@minus.07em}

\begin{document}

\title{Flow-Induced Dorso-Ventral Deformation Enhances Propulsive Efficiency in Flexible Caudal Fins}
% \title{Computational Investigation of Propulsive Efficiency Enhancement in Flexible Caudal Fins}
\author{Sushrut Kumar, Matthew J. McHenry, Jung-Hee Seo, Rajat Mittal}

\address{Department of Mechanical Engineering, Johns Hopkins University}
\address{Department of Ecology and Evolutionary Biology, University of California, Irvine}
\ead{mittal@jhu.edu}
\vspace{10pt}
\begin{indented}
\item[]June 2025
\end{indented}

\begin{abstract}
Fish swim with flexible fins that stand in stark contrast to the rigid propulsors of engineered vehicles. Using numerical simulations of the dynamics of flow-structure interaction (FSI), we have found that dorso-ventral deformation in flexible caudal fins results in a 70\% increase in efficiency of caudal fin swimmers compared to a rigid fin generating the same amount of thrust. By correlating fin deformation to the flow physics, we find that the greater power requirements of rigid fins can be largely attributed to their propensity to generate high-magnitude lateral forces. In contrast, flexible fins achieve high efficiency local-redirection of force where deformations orient pressure forces on the fin in fore-aft and dorso-ventral directions to reduce the power demand of generating thrust forces. These deformations occur at phases in the tail-beat cycle where the fin experiences large lateral velocities and pressure differentials and this reduces the net power expended by the flexible fins. In this way, the flexibility of a caudal fin offers a simple and elegant solution for efficient locomotion which does not require sensing, computation and control that might otherwise be provided by the nervous system of a fish or a computer within a underwater vehicle. These flow-induced dorso-ventral fin deformations therefore imbue a mechanical intelligence in these fins that provides propulsive advantages to caudal fin swimmers and they also offer solutions for efficient propulsion in engineered systems.
\end{abstract}

%\maketitle

\section{Introduction}

{N}ature excels at utilizing soft materials and structures that bend, deform, and stretch as they interact with the environment. 
In contrast to the `robotic' motion of engineered systems, animal locomotion is characterized by smooth movement that is largely facilitated by passive mechanical properties.
Examples include the wing deformation of flying insects and birds \cite{combes2003flexural, muijres2008leading, alexander1991energy}, the stability enhancement provided by limb deformation in insects \cite{dudekIsolated2007, jindrich2002dynamic} and the muscle-tendon compliance of vertebrates \cite{daley2006running, roberts2011flexible}.
These passive mechanical properties, resulting from \replaced{hundreds of millions of years}{millions of years} of evolution, endow animals with a rapid and robust capacity to respond to perturbations.
By facilitating smooth and stable motion with passive mechanics, less sensing, control and computation may be required to control these dynamic systems \cite{zhou2025effect}.
This capacity to offload computation for passive mechanics in a dynamic system is known as ``mechanical intelligence'' and it offers inspiration for the design of a variety of robots and other devices \citep{gupta2021embodied, zhaoExploring2024, pfeifer2006body, Howard2019}.
However, in the arena of biolocomotion we still have much to learn about the extent to which passive mechanical properties enhance the energetic efficiency of biological and engineered systems.

Swimming offers a compelling system to test the efficiency gains from mechanical intelligence in animals.
The passive mechanics of a swimmer's body and fins deform as a consequence of their coupled mechanical interactions with the surrounding water.
This flow-structure interaction (FSI) is widely appreciated to play a major role in the dynamics of swimming animals such as fish, cetaceans, and pinnipeds.
These animals use body and appendage  (i.e., a fin, flipper, or fluke) deformation to propel themselves through water \citep{fish2006passive, Long1996b}. 
For example, FSI dynamics enable the body of trout to swim in a wake with minimal muscle activity to realize an energetic benefit \citep{Taguchi2011a, Liao2004b}. 
Under these conditions, it is even possible for the body of a \emph{dead} fish to swim momentarily by undulating through a turbulent wake \citep{beal2006passive,Liao2003f}.
Despite broad appreciation for the role of FSI in swimming, the mechanisms of efficiency gains remain mysterious.

The current study explores propulsion with computational fluid dynamic modeling of fish swimming.
We focus on the caudal fin, which is the primary organ for thrust generation in a diversity of species \citep{lighthill1975mathematical,seo2022improved,zhou2025hydrodynamically, zhou2024effect}.
Caudal fins in many teleost fish are highly flexible membrane-like structures strengthened by slender bony rays that extend from the point-of-attachment on the peduncle to the tips of the fins \citep{bainbridge1963caudal,lauder2000function}. 
In many fishes, the dorsal and ventral regions of the caudal fin have a higher density of rays \citep{bainbridge1963caudal,lauder2000function} and flexor muscles preferentially attach and actuate the dorsal and ventral margins of the caudal fin \citep{flammang2008speed}. In contrast, the central part of the fin has a lower density of rays and is therefore more flexible. As the fin is actuated at the dorsal and ventral regions, its central area lags behind due to inertial and FSI effects and this results in a dorso-ventral curvature or ``billowing'' of the fin \citep{bainbridge1963caudal}. 

Bainbridge \citep{bainbridge1963caudal} was one of the first to quantify this dorso-ventral fin billowing deformation based on films of the dace (\emph{Leuciscus leuciscus}) and he hypothesized that this deformation helps smoothen out the time-variation of thrust\deleted{, }. 
% However, he was unable to verify this hypothesis. 
This deformation has been observed in other teleost species with homocercal (i.e. symmetrical) caudal fins, including the bluegill sunfish \citep{flammang2008speed}, mackerel \citep{gibb1999tail}, and batfish (Fig. \ref{fig:SchemaModel}A). 
However, Bainbridge's ideas remain untested because kinematics alone do not permit an interrogation of the FSI dynamics that generate dorso-ventral deformation.
Previous simulation-based studies of caudal fin hydrodynamics have either focused on \emph{antero-posterior} deformation \citep{bergmann2014effect} or \emph{imposed} dorso-ventral deformation on the fin \citep{fernandez2021effect}. Caudal fin ``cupping'' has been shown to improve thrust performance in biorobotic swimming machines \cite{esposito2012robotic}, but the underlying flow physics was not elucidated. Thus, the effect of FSI-induced dorso-ventral deformation on the propulsive performance of caudal fins remains unexamined. 

In the present study, we use FSI simulations to investigate the propulsive performance of a canonical model of a flexible membranous homocercal caudal fin (Fig. \ref{fig:SchemaModel}). This fin is actuated with a flapping motion at the stiff dorsal and ventral margins, but is allowed to deform passively in the central region as a consequence of fluid forces and the flexibility of the fin (details are provided in the next section). We examine how the stiffness of the central region influences thrust generation and power expenditure.
This is achieved by modeling the caudal fin with three levels of flexibility. Fins of intermediate (Fin-IF) and high flexibility (Fin-HF) are constructed with the same elastic material, but respectively with intermediate ($h^*=0.04$, where $h^*$ is the ratio of thickness to chord length) and thin ($h^*=0.02$) membranes.
These compliant fins are compared against a perfectly rigid fin (Fin-R) that permits no material compliance. 
Our comparisons quantify the degree to which passive deformation in response to flow-structure interaction modifies the propulsive performance and energetic costs of fin motion.

\section{Computational Methods}
\subsection*{Fin Model}
The caudal fin in the current study is assumed to have a canonical trapezoidal shape (Fig. \ref{fig:SchemaModel}D). The fin is comprised of stiff rays at the dorsal and ventral edges, and these are joined together with an upstream edge of the fin, which represents the peduncle. The two rays and the peduncle form the frame of the fin and a flexible membrane capable of supporting extensional and bending stresses (technically, a ``shell'') is stretched across this stiff frame. The frame is driven with a flapping motion given by
\begin{equation}
\theta(t) = \theta_0 \cos{(2 \pi f t)}; \; \, \, 
A(t) = A_0 \sin{(2 \pi f t)}
\end{equation}
where $A(t)$ is the lateral (heaving) displacement of the fin peduncle, and $\theta(t)$ is the \replaced{yaw}{pitch} angle of the fin at the peduncle. The membrane experiences hydrodynamic forces during the fin motion, and this, combined with its inertia, drives the deformation of the fin. The key non-dimensional parameters are the \replaced{yaw}{pitch} amplitude $\theta_0$, heave amplitude $A_0/C$ (where $C$ is the chord of the fin), Reynolds number (Re=$U_\infty C/\nu$) and Strouhal number (St$=2 A_0 f/U_\infty$). In the current study, we fix the values of  $\theta_0$ and $A_0/C$ to 30$^o$, 0.4, respectively. The \replaced{yaw}{pitch} and heave amplitude are consistent with previous studies \citep{DONG_MITTAL_NAJJAR_2006,bergmann2014effect,fernandez2021effect} and are a reasonable representation of the kinematics of the caudal fin of fish. Previous studies on flapping wings and fins \citep{anderson1998oscillating,sekhar2019canonical} have demonstrated that the Strouhal number (St) dominates the flow physics among the various non-dimensional parameters. Moreover, if the Reynolds number is sufficiently high so that the boundary layers are thin compared to the overall movement of the control surface, the resulting flow physics will be a reasonable approximation to the flow at a higher Reynolds number. In the current study, we fix the Reynolds number to 1000 and perform grid-refinement studies to ensure that the flow is resolved.

\subsection{Flow Simulation}
We perform the direct numerical simulations (DNS) of the flow by solving the incompressible Navier-Stokes equation described as:
\begin{equation}
    \frac{\partial u_i}{\partial x_i} = 0; \, \,
    \frac{\partial u_i}{\partial t} + \frac{\partial u_j u_i}{\partial x_j} = -\frac{\partial p}{\partial x_i} + \frac{1}{\text{Re}} \frac{\partial^2 u_i}{\partial x_j x_j}
    \label{eq:NS}
\end{equation}
where $u_i$ is the velocity and $p$ is the pressure. The governing equations are \deleted{discredited. } discretized in space and time using the $2^\text{nd}$-order finite-difference schemes and integrated in time using the operator splitting fractional step method. Here, the Eqn. \ref{eq:NS} is split into the advection-diffusion (ADE) and pressure Poisson equation (PPE), and this results in two linear systems connected through an intermediate velocity. The linear system for ADE is solved using a line successive-over-relaxation solver, and the system for PPE is solved using a stabilized biconjugate-gradient solver. We utilize our in-house numerical solver ViCar3D \citep{mittal2008versatile, mittal2025freeman} for these simulations. The flow solver has been validated for a variety of fluid dynamics studies, including coupled FSI simulations \citep{shoele2014computational,shoele2016flutter,bailoor2021computational}, fish fins \citep{DONG_BOZKURTTAS_MITTAL_MADDEN_LAUDER_2010}, and flapping insect wings \citep{zheng2013multi}. The placement of the fin within the computational domain of size $26C\times 40C\times 40C$\added{.}
%is shown in figure \ref{fig:compSetup}(a). 
A uniform grid is employed in a cuboidal region around the body to resolve the boundary layers and vortex structures, and the grid is stretched away outside this region to the outer boundary. We employ a grid of $256\times 256 \times 240$ (about 15 million points), with about 100 points along the fin chord length. We have performed extensive grid refinement studies to ensure the simulations are grid-converged (See appendix \ref{app:grid_indep}). 
Each simulation is carried out for three flapping cycles, and the results for the 3rd cycle are used for all the analysis \added{(see appendix B)}. 

\subsection{Fin Membrane Dynamics Model and FSI Coupling}
We utilize a connected spring-mass network model to simulate the hydro-elastic deformation in the elastic membrane. This class of model offers several computational advantages in studies with a focus on fluid dynamics. Moreover, the model is simple to couple with our sharp interface immersed boundary method. The fin surface is described by a mesh with triangular elements with the triangle vertices serving as ``nodes'' for the discretization of the dynamical equations. The following ordinary differential equation \eqref{bigode} governs the structural dynamics. 
\begin{equation}
    \label{bigode}
    M \frac{d^2 \mathbf{X}(t)}{dt^2} = \mathbf{F}_\text{hydro}-\mathbf{F}_\text{elas} - \zeta\frac{d \ \mathbf{X}(t)}{d t},
\end{equation}
In Eqn. \ref{bigode}, $M\in \mathbb{R}^{N\times N}$ is the diagonal matrix with nodal masses, $\mathbf{X}(t)\in \mathbb{R}^{N\times3}$ contains the node positions in 3D space. $\mathbf{F}_\text{hydro}$ and $\mathbf{F}_\text{elas} \in \mathbb{R}^{N\times3}$ are the hydrodynamic and elastic forces applied on the nodes, respectively. Material damping is modeled using the last term in Eqn. \ref{bigode}, with $\zeta$ being the structural damping coefficient. This method has been employed previously to model fluid-structural interaction of shells in a variety of applications \citep{marco2016moving,fedosov2010systematic,spandan2017parallel} and we present a detailed description along with the validation of the model in our earlier work \citep{kumar2025computational}. 
%A key challenge in simulating systems involving low-density ratio using partitioned method like ours, is the presence of added-mass instabilities. Techniques like implicit coupling is popularly used to eliminate the instabilities but is computationally expensive and doesn't always guarantee convergence. Hence, we utilize the force partitioning method \citep{menon2022contribution} to estimate the added-mass force and stabilize the numerical instabilities. 

\begin{figure*}[th!]
    \centering
    \includegraphics[width=0.8\textwidth]{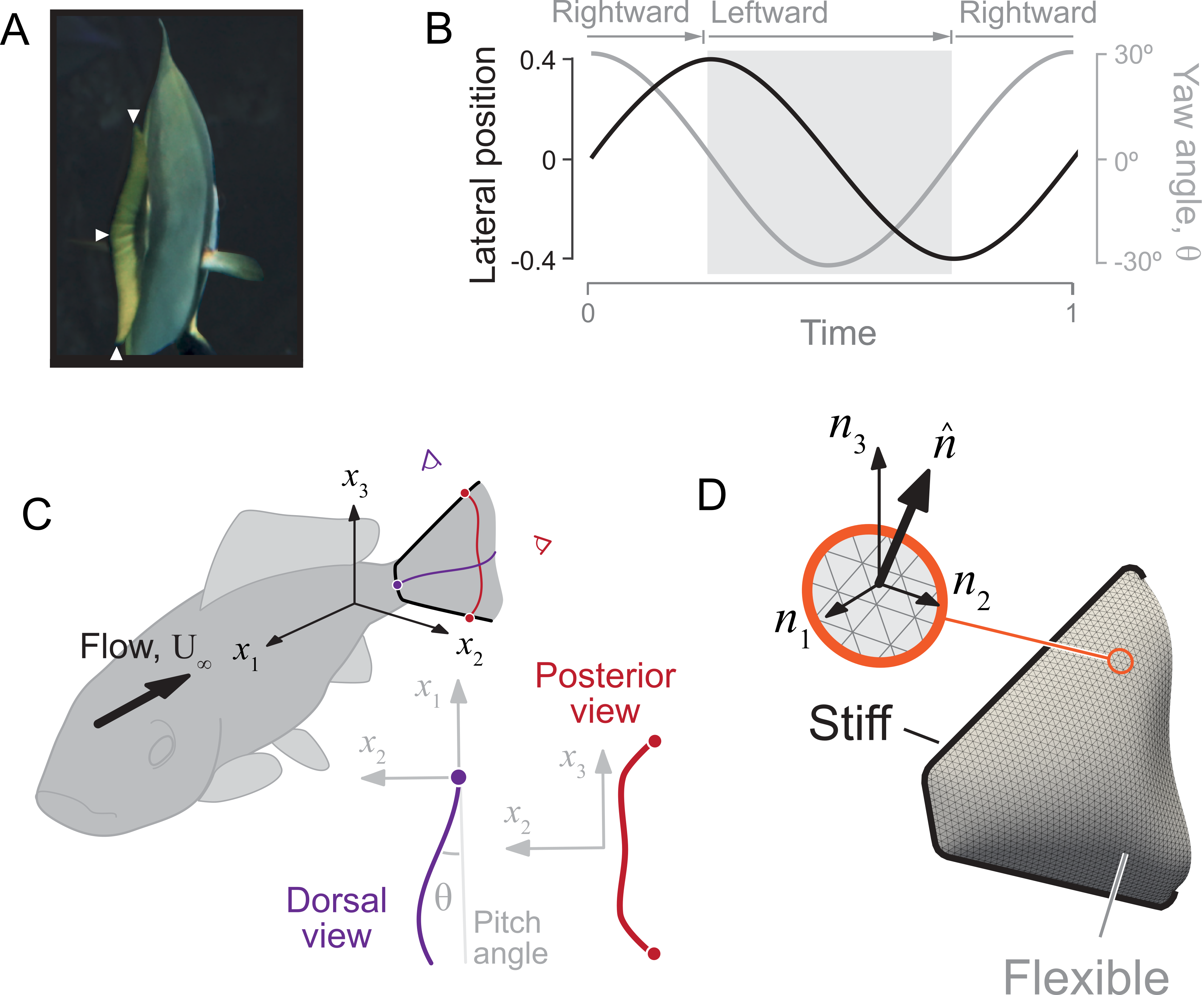}
    \caption{Modeling the flow-structure interactions of caudal fins.
    (A) Video recording of a batfish (\textit{Platax orbicularis}) from a posterior view shows how the mid-chord of the fin bends (right arrow) relative to the dorsal (down arrow) and ventral (up arrow) margins of the caudal fin as it beats toward the right.
    (B) Model fins are actuated at the leading margin (at the caudal peduncle) of the fin with sinusoidal changes in lateral position (along \textit{x}$_2$, in black) and in \replaced{yaw}{pitch} angle (gray curve) over normalized time ($t/T$, where $T$ is the tail-beat period).
    (C) \replaced{Schematic of a fish for visual context of the flexible fin operating in the Cartesian coordinate(with axes \textit{x}$_1$, \textit{x}$_2$, and \textit{x}$_3$) with deflection observed along mid-chord(purple line) and along a dorso-ventral span (red curve). In this work, only the caudal fin is simulated and not the body.}{Our model of the caudal fin operates in a Cartesian coordinate system (with axes \textit{x}$_1$, \textit{x}$_2$, and \textit{x}$_3$) with deflections that may be observed along (C) the mid-chord (blue curve) in 2D from a dorsal view and along a dorso-ventral span (red curve) from a posterior view.}
    (D) The FSI model included an elastic mesh between rigid dorsal and ventral margins (black curves), with deflections determined by the material stiffness and fluid forces.
    The inset shows a detail of the mesh, with the normal direction ($\hat{n}$) of a single triangle highlighted, with its components (\textit{n}$_1$, \textit{n}$_2$, and \textit{n}$_3$) shown within the coordinate system.
    }
    \label{fig:SchemaModel} %Qisovort
\end{figure*}
While the fin is geometrically represented as a zero-thickness membrane in the flow solver, the dynamical model of the membrane employs a finite thickness given in the above equations by $h^*=h/C$. This parameter appears in both the elastic and bending \replaced{spring constants}{moduli} and is therefore, a convenient parameter for controlling the flexibility of the fin membrane. The non-dimensional material properties of the fin membrane are based nominally on the properties of animal tissue as follows: membrane (solid) to fluid density ratio $\rho^*(={\rho_s}/{\rho_f}) = 1$; Elastic Modulus $E^* (={E}/{\rho U_\infty^2}) = 10^6$; and the Poisson's ratio $\nu=0.4$. The damping coefficient \replaced{$\zeta$}{C} is unknown, and we chose an intermediate value of $\zeta(=C/2\sqrt{k_b m})=1$) which prevents excessive numerical oscillations in the membrane. Tests show that the overall results regarding thrust, power, and efficiency are relatively insensitive to moderate variations in these properties.

\subsection{Hydrodynamic Forces and Power Expenditure}
The power expenditure of the fin is primarily associated with the work done by the fin against hydrodynamic forces. The hydrodynamic forces can be expressed as $\mathbf{F_\text{hydro}} = \left( \Delta p \, \hat{n}, \Delta \tau \, \hat{t} \right) dS $, where $\Delta p$ and $\Delta \tau$ are the pressure and shear stress differentials across the fin surface, $\hat{n}$ and $\hat{t}$ are the unit vectors normal and tangential to the fin surface, and $dS$ is an elemental area on the fin surface.

The thrust and lateral forces on the fin are obtained by integrating the $x_1$ and $x_2$ components of 
$\mathbf{F_\text{hydro}}$ over the fin surface. The power expended by the fin is dominated by the work done against the hydrodynamic forces and is given by $P=\int_{S_\text{fin}} \mathbf{F_\text{hydro}}.\mathrm{V}_\text{fin} \, dS$. 
% Power requirements were approximated by considering the forces and moments generated by the pressure distribution along the fin.
% While our simulations considered viscous forces in the fluid, the forces upon the fin are dominated by the pressured generated by flow. 
% The power expended by the fin generated by these pressure can be approximated very well by
%In order to understand how local variations of pressure on the surface can affect the thrust performance, we consider the expression for power shown earlier. In this expression, the velocity of an elemental area on the fin is given by $\mathrm{V} (X_1,X_2) = \left( (\dot{A} \cos{\theta}-X_1 \dot{\theta}) \, \hat{n}, (-\dot{A} \sin{\theta}) \, \hat{t}, 0 \right)$, where $(X_1,X_2)$ are the chordwise and vertical coordinates attached to the fin with the origin at the center of the peduncle. 
We use numerical integration over the fin surface to obtain the integral values associated with the hydrodynamic forces and power.

\section{Results}
\begin{figure*}[th!]
    \centering
    \includegraphics[width=1\textwidth]{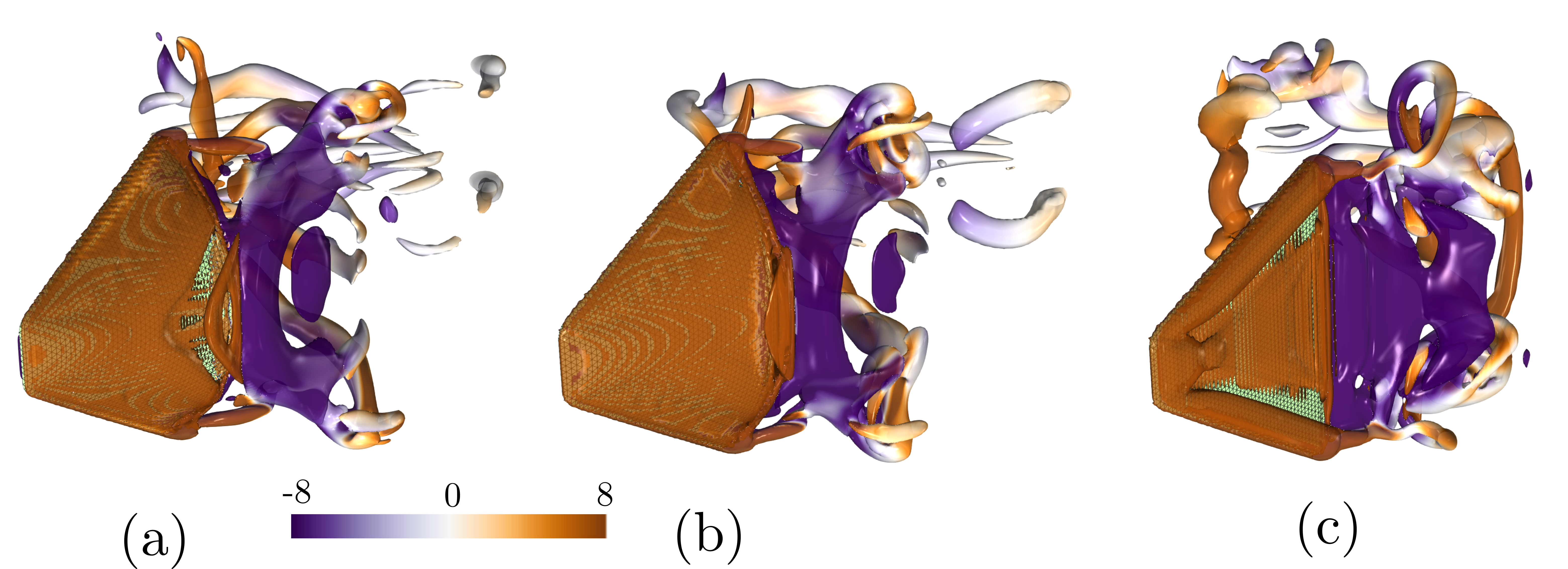}
    \caption{Isosurface of $Q$-Criterion colored with spanwise vorticity \added{at $t/T=0.125$}. (a) Fin-HF, (b) Fin-IF, (c) Fin-R}
    \label{fig:FlowStructures} %QIsoVort (E-H)
\end{figure*}

\begin{figure*}[th!]
    \centering
    \includegraphics[width=0.9\textwidth]{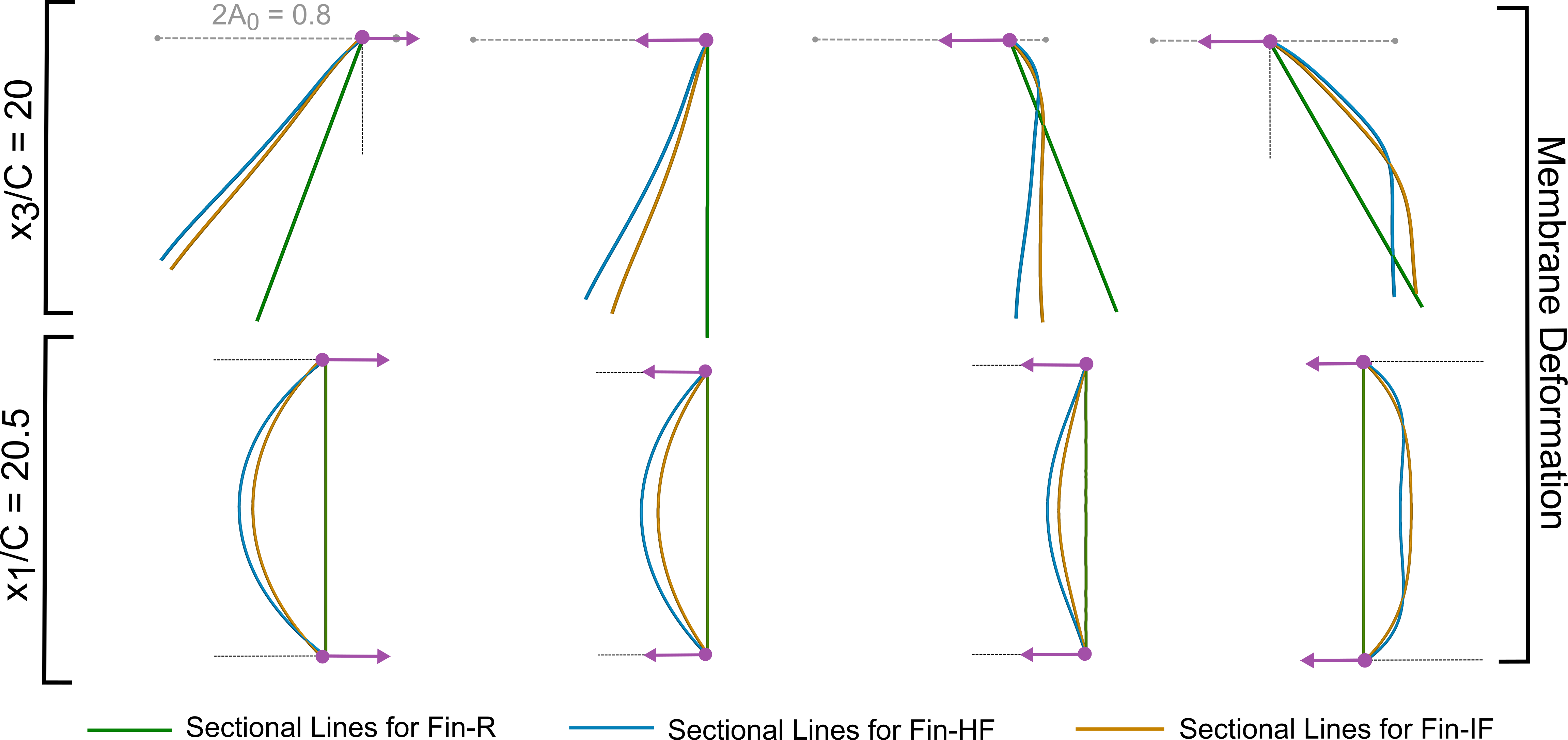}
    \caption{Sectional lines showing cross-sectional deformation in the fish caudal fin}
    \label{fig:deformation} %QIsoVort (E-H)
\end{figure*}

We performed a series of FSI simulations over a range of Strouhal numbers ($0.3 \leq \text{St} \leq 0.5$) to evaluate fin performance. 
The Strouhal number ($\text{St} = 2A_of/U_\infty$, where $f$ is fin flapping frequency, $2A_o$ is the peak-to-peak fin amplitude, and $U_\infty$ is the swimming speed) indicates the velocity of the fin relative to the oncoming flow. \added{The Strouhal number was varied by changing the flapping frequency while keeping the incoming flow velocity constant at $U_\infty = 1$.} 
The Strouhal numbers considered here extends over a range where a variety of swimmers and fliers have been shown to operate during steady swimming and flying \cite{triantafyllou2000, rohr1998observations,taylor2003}.  Fig. \ref{fig:FlowStructures} shows the representative vortex structures for the three fins (high-flexibility, intermediate flexibility and rigid fin). 
Fig. \ref{fig:deformation} shows four snapshots over one flapping cycle that compare the fin deformation for a rigid and a flexible (Fin-HF) at a Strouhal number of 0.3. Specifically, Fig. \ref{fig:deformation} compares the chordwise and spanwise midlines of all three fins at these same time instances during the cycle. We note that both the deformable fins exhibit very significant deformations with a large concave ``billowing'' that is prominent during mid-stroke and this is inline with experimental observations \cite{bainbridge1963caudal}. The billowing of the fin will change the angle of the incoming flow relative to the surface at the leading-edge of the ventral and dorsal edges of the fin. It also enhances the local yaw angle of the fin relative of the flow direction at all locations on the fin, but particularly so at the fin midline. As we will show later, both of these features have implication for the thrust and efficiency of the fin.

\begin{figure*}[th!]

\subsection{Propulsive Efficiency}

    \centering
    \includegraphics[width=1.0\textwidth]{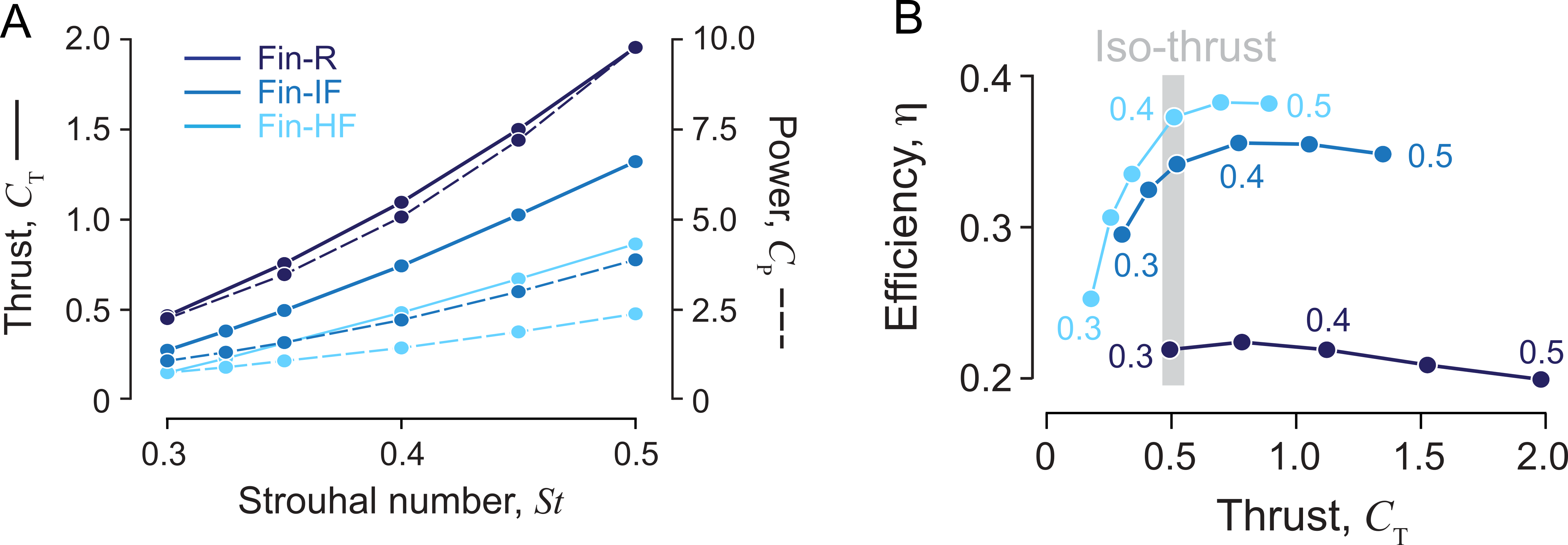}
    \caption{Metrics of hydrodynamic performance for the fins: 
    The time-averaged (A) thrust coefficient (solid curves), power coefficient (dashed curves), and (B) efficiency are shown for simulations performed at varying Strouhal numbers (values shown for $St = 0.3, 0.4,$ and $0.5$). Simulations with comparable $C_\text{T}$ values are employed for the `Iso-thrust' comparison (gray bar).
    }
    \label{fig:ThrustPowerEfficiency} %QIsoVort (E-H)
\end{figure*}
% Irrespective of the mechanical properties of a swimmers, their  efficiency varies with the Strouhal number.
Fig. \ref{fig:ThrustPowerEfficiency} shows the time-averaged thrust coefficient, power coefficient, and Froude efficiency for the 18 cases simulated in this study. Consistent with physical and analytical models of hydrofoils \citep{triantafyllou2000,triantafyllou1993optimal}, the coefficient of non-dimensional thrust and its power requirements increase with the Strouhal number in our FSI simulations (Fig. \ref{fig:ThrustPowerEfficiency}).
The increase in thrust ($\bar{C}_T=\bar{T}/(0.5 \rho U_\infty^2 S_\text{fin}$), where $\rho$ is water density, $U_\infty$ is flow speed, $T$ is thrust and $S_\text{fin}$ is the area of the fin) reflects the greater momentum that the foil imparts to the fluid as it increases its frequency and/or amplitude at higher Strouhal numbers.
However, the power required to generate that thrust ($\bar{C}_P=\bar{P}/(0.5 \rho U_\infty^3 S_\text{fin}$, where $P$ is power) varies at a different rate (Fig. \ref{fig:ThrustPowerEfficiency}), which has implications for the efficiency of propulsion ($\eta=\bar{T}U_\infty/\bar{P}$).

The thrust and the power requirements among fins of different flexibility yield more complex trends in propulsive efficiency.
The maximum efficiency of the intermediate fin (Fin-IF) is 58\% greater than that of the rigid fin (Fin-R), whereas for the most flexible fin (Fin-HF, Fig. \ref{fig:ThrustPowerEfficiency}), this advantage stands at 70\%. Over the entire range of Strouhal numbers examined, the rigid fin is substantially outperformed in terms of efficiency by the flexible fins and this performance increases with increasing fin compliance.  
Interestingly, the figure shows that the maximum efficiency occurs at a similar level of thrust (at $C_T \sim 0.75$) for all the fins. 
We, therefore, focus our analysis of the FSI mechanisms on the individual simulations for each fin that attains close to maximum efficiency. Within the context of this `iso-thrust' comparison, the efficiency of different fins may be understood by focusing on the power requirements to actuate each fin.  

Given that at these high Reynolds numbers, $\Delta p \gg \Delta \tau$, (where $\Delta p$ and $ \Delta \tau$ are the pressure and shear stress differences across the fin) the power expended by the fin can be approximated reasonably well by the work done against the pressure-induced hydrodynamic forces on the fin i.e. 
\begin{equation}
P(t) \approx  \int_{S_\text{fin}} \Delta p \, \hat{n} \cdot \vec{V} \, \text{dS},
\label{eq:PressurePower}
\end{equation}
where $V$ is the velocity of the fin.
We first examine the rigid fin to understand the dominant sources of power expenditure. 
For a flat (undeformable fin) undergoing yawing and heaving, the velocity of an elemental area on the fin is given by $\vec{V} (X_1,X_2) = \left( (\dot{A} \cos{\theta}-X_1 \dot{\theta}) \, \hat{n}, (-\dot{A} \sin{\theta}) \, \hat{t}, 0 \right)$, where $(X_1,X_2)$ are the chordwise and vertical coordinates attached to the fin with the origin at the center of the peduncle. Given that at these high Reynolds numbers, $\Delta p \gg \Delta \tau$, and $\cos{\theta} > \sin{\theta}$ for the yaw angles for these fins, the power expended by the fin can be approximated very well by
\begin{equation}
    P(t) \approx -\int_{S_\text{fin}} \Delta p \left(  \dot{A} \cos{\theta} - X_1 \dot{\theta} \right) dS
    = - \dot{A} L  + \dot{\theta} M
\label{eq:power-estimate-rigid}
\end{equation}
where $L$ and $M$ are the pressure components of the lateral force and yaw moment on the fin, respectively. We note that $L$ depends on the surface-averaged magnitude of the pressure differential but not on its distribution on the fin. On the other hand, due to moment arm $X_1$ in its expression, $M$ depends on the distribution of pressure differential.

The deformable fin presents a more complicated situation for the analysis of power because the shape of the fin deviates substantially from the flat shape to which the power estimate in Eq. \ref{eq:power-estimate-rigid} can be applied. However, the midline deformation of the deformable fin in Fig. \ref{fig:deformation} suggests that the flexible fin could be considered as a superposition of a flat fin with a yaw angle $\theta_f(t)$ equal to the average chordwise inclination of the flexible fin and a perturbation from this flat fin. With this approach, the power for the flexible yawing fin can be expressed as 
\begin{equation}
    P(t) \approx  - \dot{A}_f L  + \dot{\theta}_f M + P_D \,,
\label{eq:power-estimate-flexible}
\end{equation}
where $\dot{A}_f$ and $\dot{\theta}_f$ are the average heaving velocity and average yaw rates of the flexible fin, and $P_D$ is the residual power associated with the deformation of the fin about its chordwise mean configuration, ie. due to the curvature of the fin. 
The decomposition in Eqs. \ref{eq:power-estimate-rigid} and \ref{eq:power-estimate-flexible} for the rigid and flexible fins now allow us to explore various mechanisms that could explain the reduced power consumption of the flexible fins per unit thrust. These are detailed in the following sections.

\begin{figure*}[th]
    \centering
    \begin{minipage}[t]{0.48\textwidth}
        \centering
        \includegraphics[width=\linewidth]{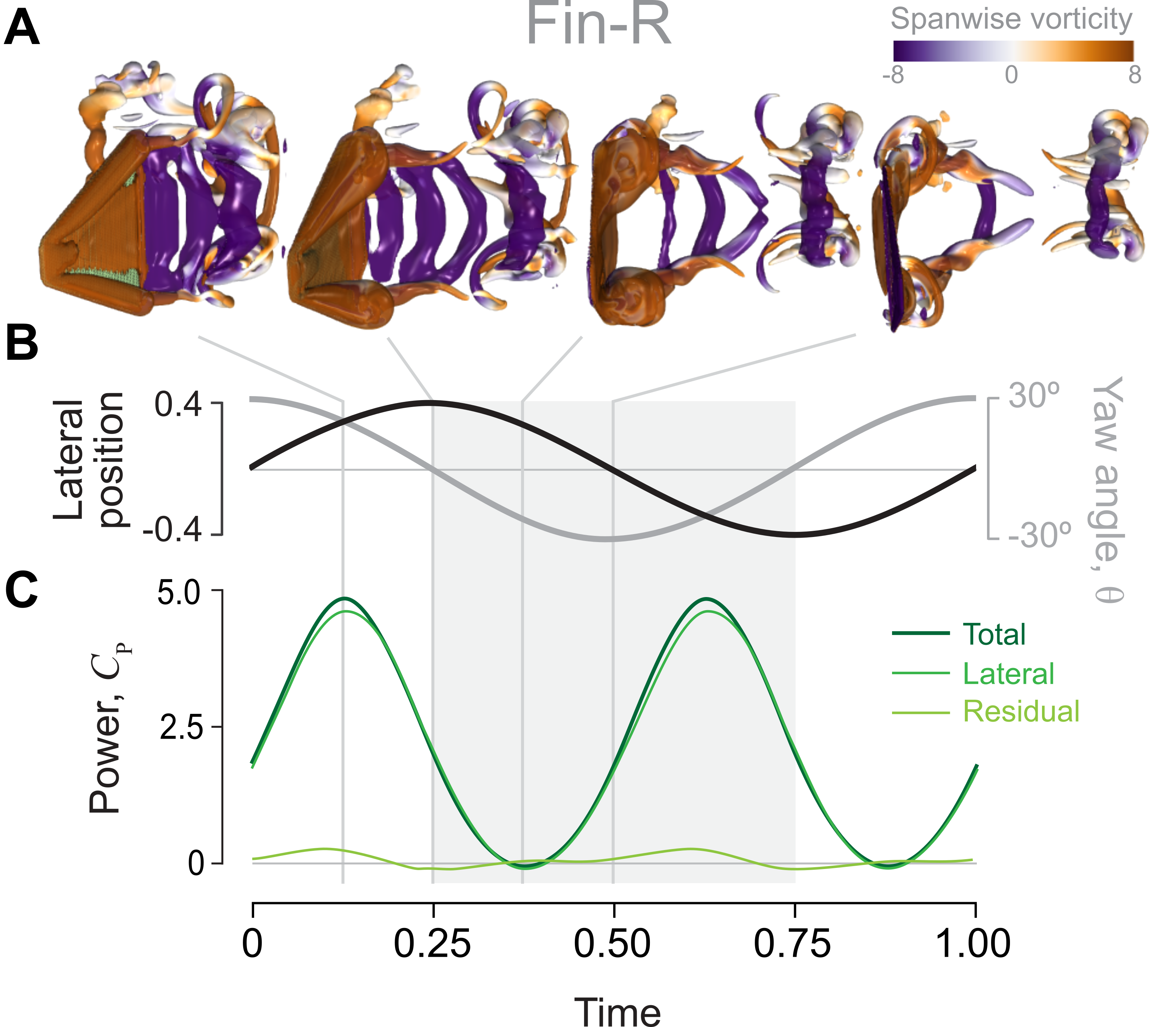}
        \vspace{2pt}
        \textbf{(i)} Rigid fin.
    \end{minipage}
    \hfill
    \begin{minipage}[t]{0.48\textwidth}
        \centering
        \includegraphics[width=\linewidth]{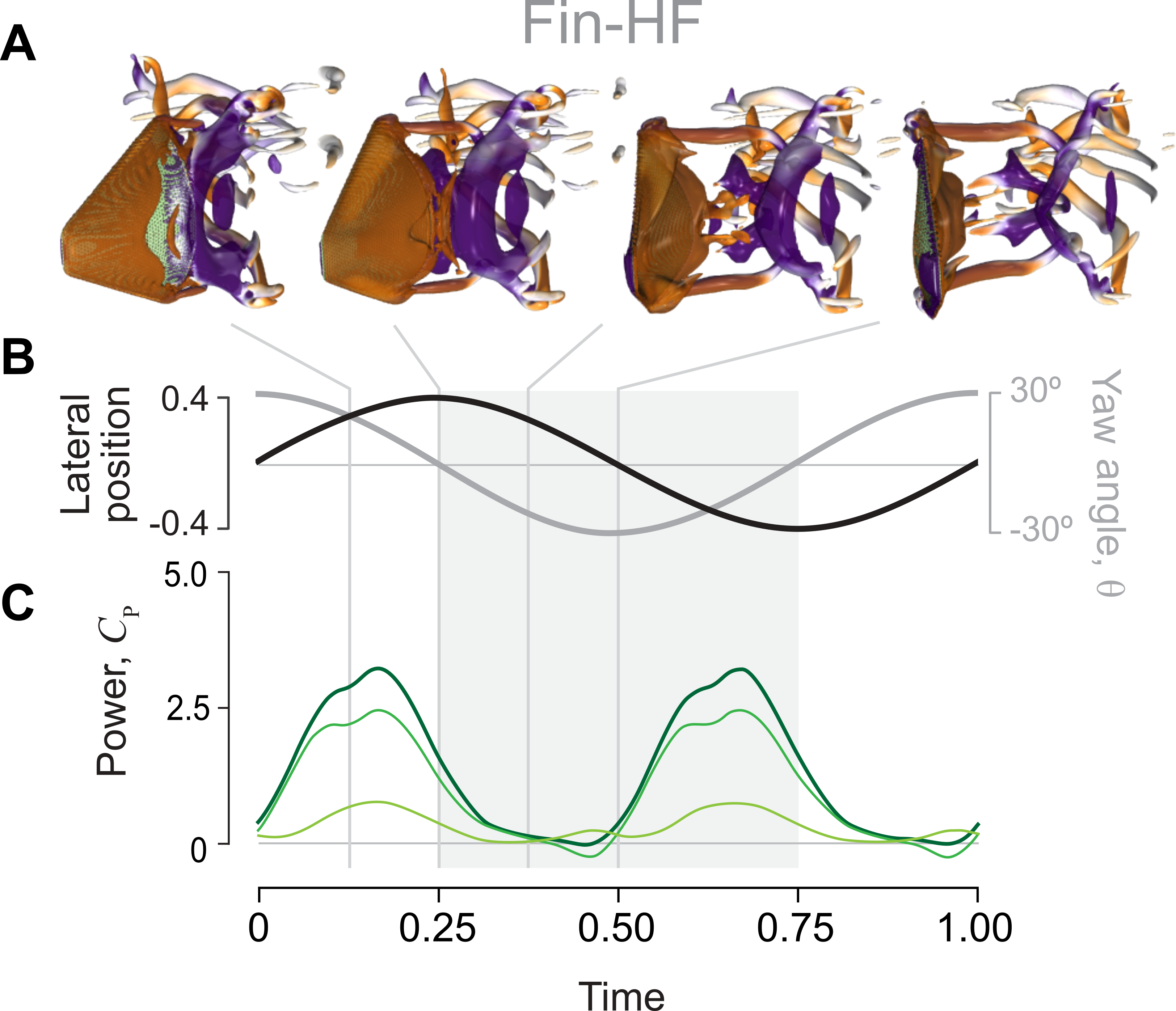}
        \vspace{2pt}
        \textbf{(ii)} Flexible fin.
    \end{minipage}

    \caption{
    Contribution of lateral forces to the power requirements for propulsion in 
    (\textbf{i}) a rigid fin and (\textbf{ii}) a flexible fin, evaluated using identical kinematics and flow conditions.
    For each case, (A) the computed flow field is visualized using isosurfaces of the spanwise vorticity
    ($Q=5$), colored by the normalized vorticity $\omega_3 C/U_\infty$.
    (B) The fins are actuated with oscillatory changes in lateral position ($x_2/C$) and yaw angle, with a representative leftward half tail-beat highlighted in light gray.
    (C) The resulting total power requirements are decomposed into the contribution associated with lateral forces and a residual term accounting for all remaining sources of power.
    }
    \label{fig:finComparison}
\end{figure*}

The high efficiency of flexible fins has the potential to be assisted by how the flow-structure interaction affects the pressure distribution along the fin. In particular, if large pressure differentials are positioned further forward (towards the peduncle), then they would tend to reduce the \replaced{yaw}{pitch}ing moment on the fin, ie. the second term on the right-hand side of Eqs. \ref{eq:power-estimate-rigid} and \ref{eq:power-estimate-flexible}. This reduced \replaced{yaw}{pitch}ing moment would reduce the power requirement for propulsion and yield a higher efficiency. Such \emph{\replaced{yaw}{pitch}-moment reduction} appears possible, given the qualitative differences in flow across fins of different flexibility (see Fig. \ref{fig:FlowStructures}).  In particular, the rigid fin develops a spiral leading-edge vortex on the dorsal and ventral edges that grows rapidly during the stroke. This fin additionally sheds several distinct vortices from the trailing-edge during each half of the flapping cycle. In contrast, the \replaced{leading-edge vortices (LEVs)}{LEVs} over the most flexible fin are more uniform in strength along the dorsal and ventral edges, and remain attached through each half of the flapping cycle (Fig. \ref{fig:finComparison}(ii)A). The flexible fin also sheds one large dorso-ventral vortex from the trailing-edge in each half of the flapping cycle, whereas the rigid fins sheds a series of vortex structures.

For the flexible fin, it is difficult to separate the effect of yaw-moment associated power from the residual power due to deformation induced curvature ($P_D$). Thus, our approach is to compute the lateral-force associated power (i.e. the first term on the right-hand-side of Eqs. \ref{eq:power-estimate-rigid} and \ref{eq:power-estimate-flexible} which is designated as $P_L$). The subtraction of this term from the total power gives us a measure of the yaw-moment associated power. 

$P_L$ is computed for all the fins by multiplying the surface-averaged lateral force on the fin with the surface averaged lateral velocity as follows:  
\begin{equation}
P_L =  \int_{S_\text{fin}} \Delta p \, \hat{n} \cdot \hat{i}_2 \,dS \times {{S^{-1}_\text{fin}}}\int_{S_\text{fin}} \vec{V} \cdot \hat{i}_2 \,dS
\label{eq:power-lateral-force}
\end{equation}
Figs. \ref{fig:finComparison}(i)C, \ref{fig:finComparison}(ii)C show the decomposition of the total power into that due to the lateral forces. The remaining portion, also plotted, is associated mostly with the yawing moment for the rigid fin and a combination of yawing moment and power associated with curvature due to deformation in the flexible fin. These plots show that the work done by the fin against the lateral hydrodynamic forces accounts for the vast majority of the power expenditure for both the rigid and flexible fin.  However, the proportions differ with fin flexibility. For the rigid fin, the work against the lateral force comprises about 97\% of the required power, whereas this is 77\% in the most flexible fin. The manner in which a flexible fin maintains attached LEVs (Fig.\ref{fig:finComparison}(ii)A) and the other changes in the flow structures cause a greater proportion of power to be applied to work against the \replaced{yaw}{pitch}ing moments as well as the power associated with the curvature induced effects of the flexible fin. However, all of the effects associated with yawing moments and curvature on the flexible fin 
tend to \emph{increase} the power expenditure and \emph{reduce} the efficiency, and therefore cannot explain increased efficiency of the flexible fins.   

The above analysis also confirms that the work done against the lateral forces is indeed the dominant contributor to power. In fact it is clear from Figs. \ref{fig:finComparison}(i)C, \ref{fig:finComparison}(ii) that the flexible fins expend much lower power than the rigid fin to generate the same total thrust. Therefore, the explanation for the higher efficiency lies with the costs of generating lateral forces and we explore this issue below.

The total work done against lateral force in one cycle is equal to 
\begin{equation}
\int_0^T P_L (t) dt = \int_0^T \dot{A}(t) L(t) dt = \int_0^T \dot{A}(t) \frac{T(t)}{ \tan{\theta(t)}} dt
\label{eq:PL}
\end{equation}
where $T(t)$ is the pressure thrust force and $\theta(t)$ is the yaw inclination of the fin. Thus for the same total thrust, the lower power could come from two possible mechanisms: the first, which we term as the  \textit{force-velocity phase mismatch} mechanism posits that power requirements are reduced for the deformable fin because of out-of-phase shift in the time-variation of the lateral force relative to the lateral velocity. This could happen for instance due to the shift in phase of the yaw-inclination associated $\tan{\theta(t)}$ term for the flexible fin. The other possibility is that the pressure force on the surface of the deformed fins is redirected by the change in average yaw inclination so as to reduce the lateral forces on the fin for a given thrust. This would manifest via an increase in the $\tan{\theta}$ term in the above expression. We refer to this mechanism as the  \textit{local-force-redirection} mechanism.
We now consider the relative contribution of each of these three mechanisms towards enhancing the efficiency of propulsion of the flexible fins (Fig. \ref{fig:ThrustPowerEfficiency}).

% \section*{Efficiency Enhancement with Flexible Fins}
\subsection{The Force-Velocity Phase Mismatch Mechanism}
\begin{figure*}[t]
    \centering
    {\includegraphics[width=0.7\textwidth]{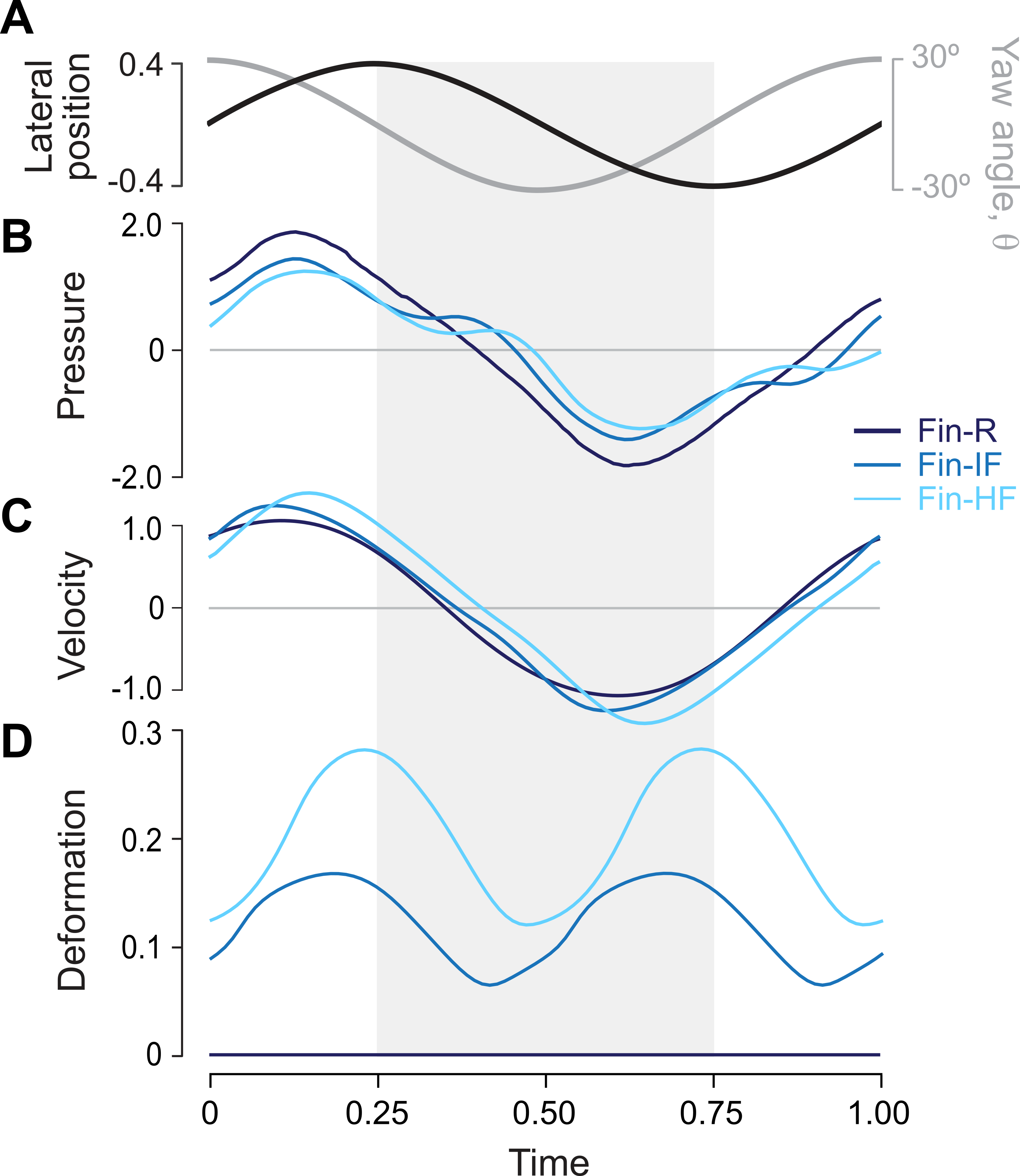}}
    \caption{Phase relationship between pressure and velocity. 
    (A) All fins were actuated with oscillations in lateral position (black) and \replaced{yaw}{pitch} angle (dark gray), with a leftward half tail-beat (gray field) highlighted.
    The resulting integrated (B) pressure differences ($1/S_\text{fin}\int \Delta p dS$) and (C) velocity ($1/S_\text{fin}\int_{S_\text{fin}} \bf{V}_\text{fin} dS$) of elements of the fin mesh (Fig. \ref{fig:SchemaModel}E) for fins that are rigid (Fin-R, dark blue), and of intermediate (Fin-IR, medium blue), and high (Fin-HR) flexibility.
    (D) Fin deformation, defined by the integrated difference in the direction of surface normals from the rigid fin ($1/S_\text{fin} \int_{S_\text{fin}} (1 - \hat{n}_\text{r} \cdot \hat{n}_\text{f})dS$ where $\hat{n}_r$ and $\hat{n}_f$ are the surface normals of the rigid and flexible fins, respectively.).
    }
    \label{fig:phase}
\end{figure*}

% In the previous section, we provided quantitative proof for the dominant role of lateral forces in the power expenditure of the fin.
The \emph{force-velocity phase mismatch} mechanism considers the possibility that flexible fins generate efficient propulsion by shifting the timing between propulsive forces and fin velocity.
Because power emerges from the product of lateral force and lateral velocity (Eqn. \ref{eq:PL}), power-efficient motion could be attained by generating high force at phases in the tail-beat cycle that correspond to low fin velocity, or vice-versa. In an extreme example, no power would be required if the lateral force and the lateral velocity were completely out of phase with each other.

Our simulations however show (see Fig. \ref{fig:phase} that for all cases, the fin moves with a velocity that is nearly in phase with the pressure differentials that generate propulsion. This can be seen in the variation of the surface-averaged pressure-differential across the fin over the flapping cycle and the speed of the fin surface, among all fins (Fig. \ref{fig:phase}B--C). We note that in addition to exhibiting similar timing, these two quantities show non-sinusoidal variations in the pressure differential for the flexible fins at the points where the fin reverses direction. We have estimated the phase between the lateral force and the average lateral velocity for all three fins and this is found to be $16^\circ$, $24^\circ$ and $11^\circ$ for Fin-R, Fin-IF and Fin-HF, respectively.
Thus, we find no consistent trend in the phase differences between force and velocity that could explain the increase in efficiency of the flexible fins. In fact, the most flexible fin (Fin-HF) that has the highest efficiency, has a phase-difference that is \emph{lower} than the less efficient rigid fin (Fin-R), and this would tend to \emph{increase} the power expenditure.
Thus, this mechanism, which is based on the phase-difference between force and velocity cannot explain the higher efficiencies of the deformable fins (Fig. \ref{fig:ThrustPowerEfficiency}B).

\subsection{The Local-Force-Redirection Mechanism}
\begin{figure*}[t]
    \centering
    {\includegraphics[width=0.6\textwidth]{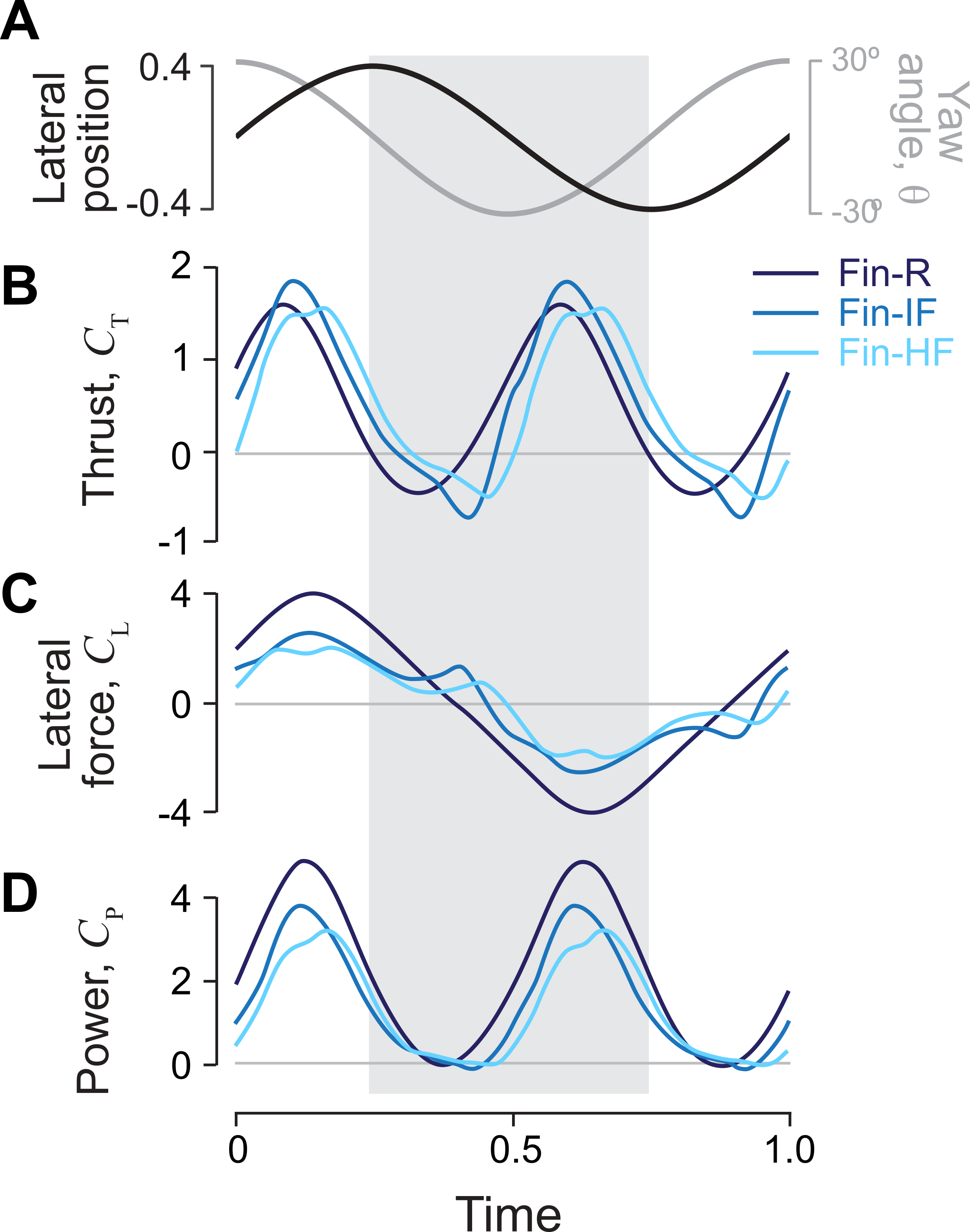}}
    \caption{Force, power, and deflection among fins are varying flexibility.
     (A) All fins were actuated with oscillations in lateral position (\added{$x_2/C$}) and \replaced{yaw}{pitch} angle (\added{$\theta$}), with a leftward half tail-beat (gray field) highlighted.
    The resulting coefficients of (B) thrust, (C) lateral force, and (D) power are shown for a rigid fin (Fin-R, dark blue), and fins of intermediate (Fin-IR, medium blue), and high (Fin-HR) flexibility.
    }
    \label{fig:flexibility_comparison} %time_series (a-d)
\end{figure*}

The difference in efficiency between the fins is apparent from an examination of the time course of the forces and power expenditure for each fin shown in Fig. \ref{fig:ThrustPowerEfficiency}.
Due to our iso-thrust condition of our comparison, the thrust for each fin is comparable across time (Fig. \ref{fig:flexibility_comparison}B).
However, the lateral forces are elevated in the rigid fin compared to the flexible fins, with a maximum force that is 153\% greater than the most flexible fin (Fig. \ref{fig:flexibility_comparison}C).
These large lateral forces do not assist propulsion but are key contributors to the power expenditure (Eqn. \ref{eq:PL}), and it is therefore unsurprising that the average power is up to 63\% greater for the rigid fin, as compared with the most flexible fin (Fig. \ref{fig:flexibility_comparison}D).
Therefore, having eliminated other mechanisms from contention, it is not surprising that the inferior efficiency of the rigid fin is found to be connected to its propensity to generate larger magnitudes of energetically costly lateral forces.

\begin{figure*}[t]
    \centering
    {\includegraphics[width=0.6\textwidth]{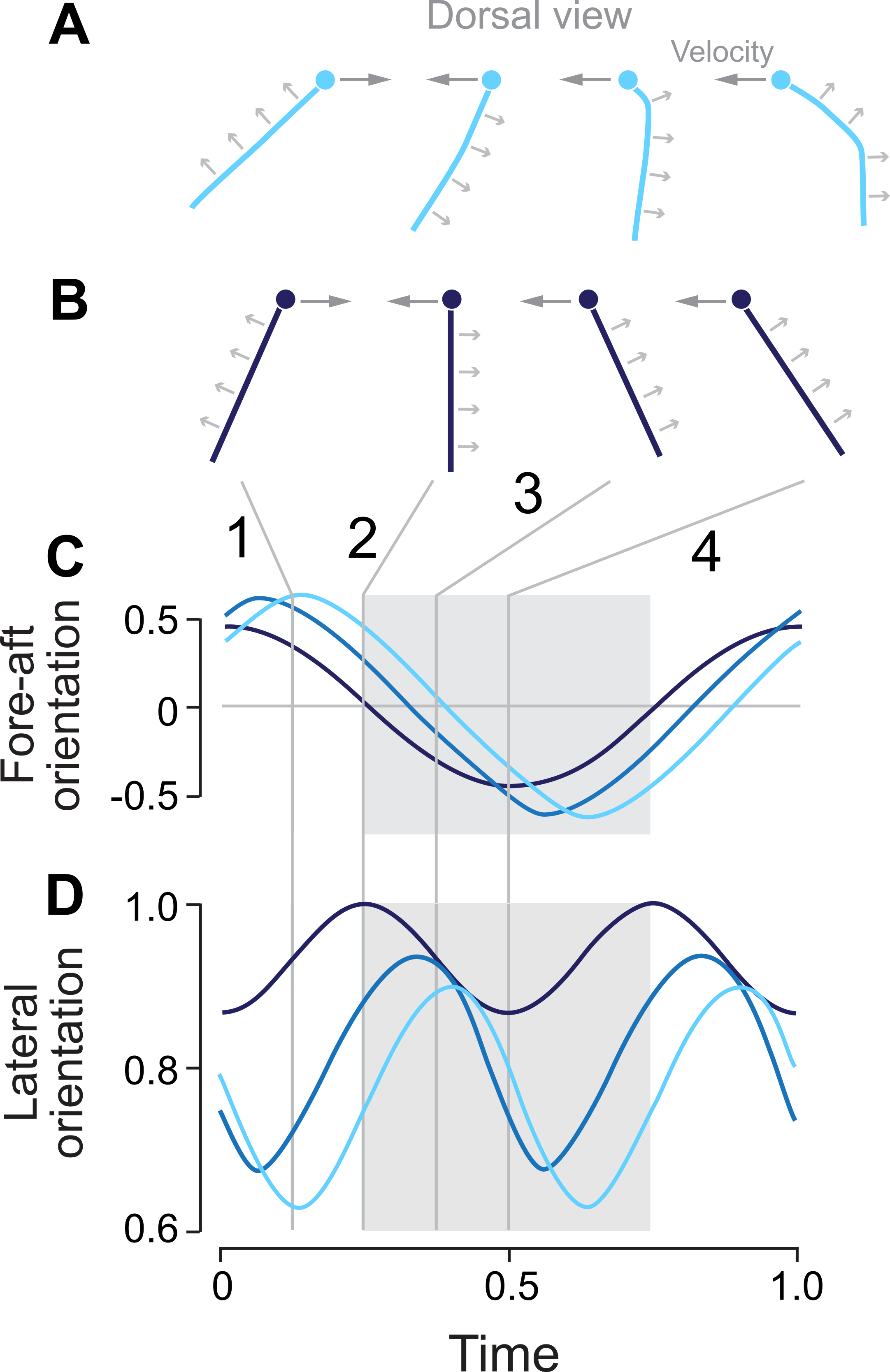}}
    \caption{The deflections of the mid-chord are shown from a dorsal view (as in Fig. \ref{fig:SchemaModel}) for the Fin-R (dark blue) and Fin-HF (light blue) models, with corresponding integrated fin area in the (C) fore-aft ($1/S_\text{fin} \int n_1 dS$) and (D) lateral ($1/S_\text{fin} \int n_2 dS$) orientations.
    }
    \label{fig:dorsal_view} %time_series (e-h)
\end{figure*}
\begin{figure*}[t]
    \centering
    {\includegraphics[width=0.6\textwidth]{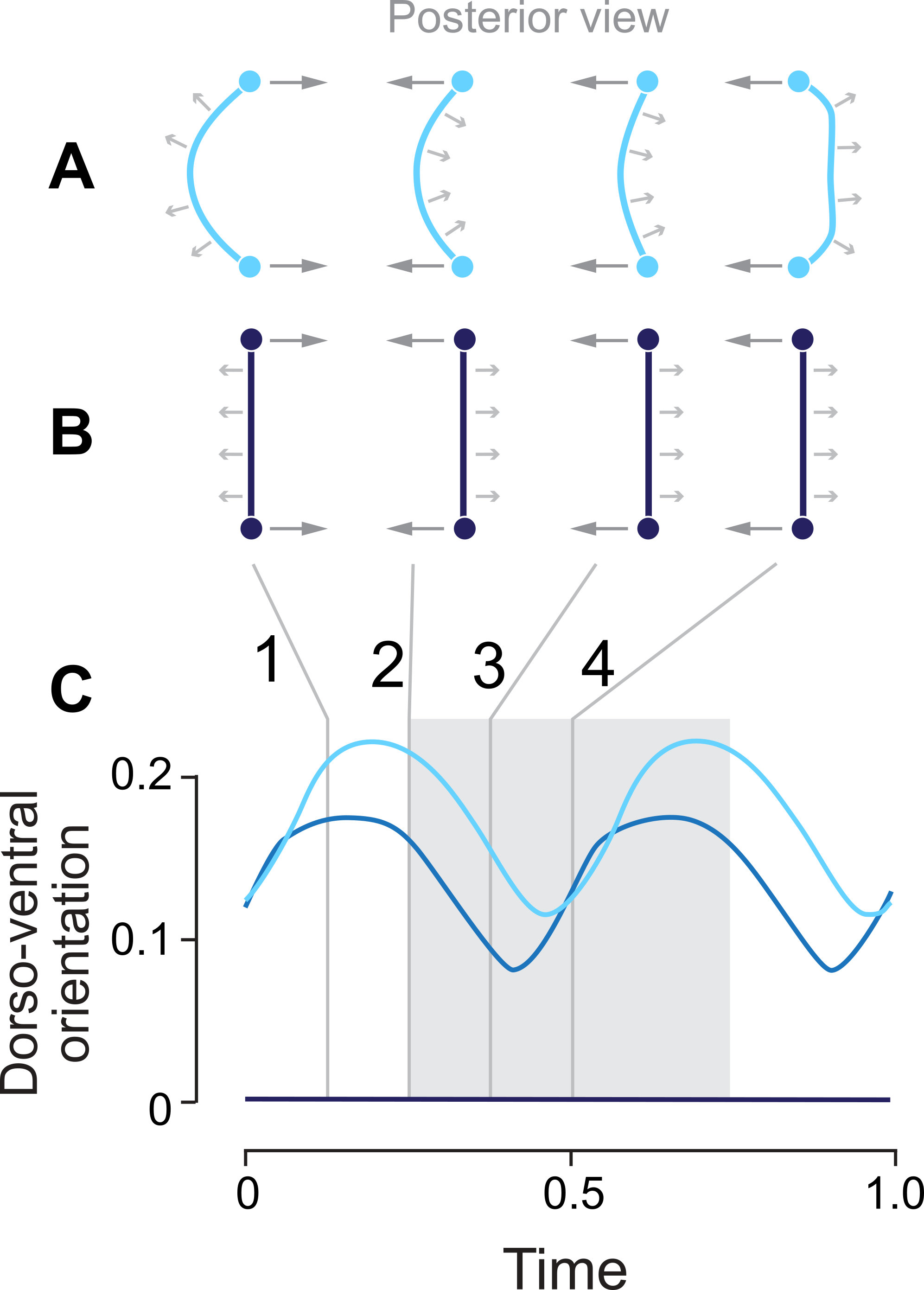}}
    \caption{
    (A-B) The fin deflect from a ventral view are shown for the models, with (C) corresponding integrated area of the fin in the dorso-ventral direction ($1/S_\text{fin} \int | n_3 | dS$). 
    }
    \label{fig:ventral_view} %time_series (i-k)
\end{figure*}
How do the flexible fins generate smaller lateral forces than the rigid fin, while still matching the thrust? The answer to this question lies in a detailed examination of the dynamic changes in the fin shape for the flexible wings (Fig. \ref{fig:dorsal_view}).
Bearing in mind that pressure acts perpendicular to surfaces (i.e. toward $\hat{n}$), we examined the orientations of the normal vectors of fin surfaces ($\hat{n}=(n_1$, $n_2$, $n_3)$) defined with respect to the coordinate system ($x_1$, $x_2$, $x_3$, Fig. \ref{fig:SchemaModel}B, E) where $x_1$ is the direction of travel of the fish (i.e. in the direction of thrust), $x_2$ is the lateral direction and $x_3$ is the dorso-ventral direction.  Averaging these components of the normal vectors across a fin's surface provides a measure of the overall orientation of the fin and hence the potential to generate lateral force or thrust.
As each fin moves through a tail-beat, the fore-aft component of its surface orientation sweeps through a positive (i.e. thrust-contributing) phase, followed by a negative period (i.e. drag-generating) before reversing direction (Fig. \ref{fig:dorsal_view}C).
This temporal pattern is similar to the oscillations in pressure (Fig. \ref{fig:phase}B) that generate the primary forces.
This pattern is similar across the fins, with a noticeably elevated orientation advantage (e.g. 27 \% greater in Fin-HF than Fin-R) in the flexible fins. This is indicative of the fact that the flexible fins are more effective in orienting pressure forces towards the thrust direction than the rigid fin.  However, a more dramatic difference is apparent when examining the lateral components of fin orientation (Fig. \ref{fig:dorsal_view}D).
All fins oscillate in this respect, as each is \replaced{yaw}{pitch}ed while being translated laterally.
However, the rigid fin maintains an elevated lateral orientation throughout each tail-beat and does not exhibit the same declines exhibited as the deflected flexible fins are swept through the fluid at relatively high speed (Fig. \ref{fig:dorsal_view}A-B).
Therefore, the fixed orientation of the rigid fin contributes to the elevated lateral forces generated through a tail-beat cycle (Fig. \ref{fig:flexibility_comparison}C).

Another factor with respect to propulsive power is the generation of forces in the dorso-ventral direction.
As a consequence of the fin deflection apparent from a posterior view (Fig. \ref{fig:ventral_view}A), the margins of a caudal fin are tilted so as to generate a vertical component to the surface normal. 
The upwards and downwards components of these forces effectively cancel each other out due to their dorso-ventral symmetry. 
However, the sum of their absolute value is substantial (Fig. \ref{fig:ventral_view}C) relative to the deformations that contribute to thrust (Fig. \ref{fig:dorsal_view}C).
These vertical forces contribute to power reduction because they redirect and further reduce the component of the pressure force in the lateral direction (Eqn. \ref{eq:power-estimate-flexible}).
In contrast, a rigid fin presents no dorso-ventral surfaces (Fig. \ref{fig:dorsal_view}J) and hence does not exploit the force redirection in dorso-ventral direction to reduce power and enhance efficiency.

The differences in the forces over the fin are therefore a combined effect of the force redirection mechanism associated with the fin geometry and the changes in the flow (and pressure) induced by fin deformation.  We have devised a method to partition these effects and to parse the relative contributions of these two effects.
This partitioning considers the local pressure differentials for elements of rigid ($\Delta p_\text{r}$) and flexible ($\Delta p_\text{f}$) fins, as well as the direction of the normal forces ($\hat{n}_\text{r}$ and $\hat{n}_\text{f}$, respectively) from those differentials.
Then, the difference of force between between the rigid and flexible fins can be expressed by the following:
\begin{align}
    \label{power_comonents}
    \begin{split}
        \underbrace{\int_{S_\text{fin}} \Delta p_\text{f} \, \hat{n}_\text{f} \, dS}_{\text{Flexible fin}} - 
        \underbrace{\int_{S_\text{fin}} \Delta p_\text{r} \, \hat{n}_\text{r} \, dS}_{\text{Rigid fin}}
        = \underbrace{\int_{S_\text{fin}} \Delta p_\text{r} \, \hat{n}' \, dS }_{\text{Geometry \, difference}}
        + \underbrace{\int_{S_\text{fin}} \Delta p' \, \hat{n}_\text{r} \, dS}_{\text{\replaced{Pressure}{Flow} \, difference}}
        + \underbrace{\int_{S_\text{fin}} \Delta p' \,  \hat{n}' \, dS }_{\text{Combined effect}}.
    \end{split}
\end{align}
where  $\Delta p' = \Delta p_\text{f} - \Delta p_\text{r}$ represents the difference in pressure differentials between the flexible and rigid fins, and $\hat{n}' = \hat{n}_\text{f} - \hat{n}_\text{r}$, is the change in surface inclination between the flexible and rigid fins. 
This equation expresses this difference in force between flexible and rigid fins in terms of the geometry only (first term), the difference due to the flow (second term), and a combined ``nonlinear'' effect of the two (third term). 

The partitioned components of thrust offer insightful metrics for propulsive differences between the flexible and rigid fins.
The flexible fins generate lower pressure differentials (Fig. \ref{fig:phase}B) for the same amount of mean thrust (Fig. \ref{fig:flexibility_comparison}B), and hence negative values for the pressure difference(e.g. -0.4 in Fin-HF, Table \ref{tab:example_table}).
This negative contribution in pressure is more than compensated for by the difference due to geometry (e.g. +0.57 in Fin-HF, Table \ref{tab:example_table}).
The lateral force, calculated for a half-cycle due to its zero cycle-average, reflects a power requirement that does not contribute to propulsion.
The most flexible fin generates only half the lateral force of the rigid fin, with geometry-difference and pressure-difference terms contributing similarly to this reduction (Table \ref{tab:example_table}). 
Thus, while the pressure-difference term diminishes both the thrust and the lateral force, the geometry-difference increases the thrust but decreases the lateral force. Thus, the change in fin geometry due to flexibility is clearly the defining factor in the high efficiency of the flexible fins.  
\begin{table}[h]
    \centering
    \renewcommand{\arraystretch}{1.2} % Adjust row spacing
    \resizebox{\columnwidth}{!}{ % Resize table to fit column width
    \begin{tabular}{lcccccc}
        \toprule
        & & Flexible & Rigid  & Geometry  & Pressure  & Combined  \\
        & &  fin &  fin &  difference &  difference &  effect \\
        Fin  & Forces & $\int \Delta p_f n_f dA$ & $\int \Delta p_r n_r dA$ & $\int \Delta p_r n' dA$ &  $\int \Delta p' n_r dA$ & $\int \Delta p' n' dA$ \\
        \midrule
        Fin-IF & $C_T$ & 0.64 & 0.64 & 0.43 & -0.30 & -0.12 \\
        & $C_L$ (1/2 cycle) & 1.28 & 1.96 & -0.40 & -0.38 & 0.10 \\
        \midrule
        Fin-HF & $C_T$ & 0.63 & 0.66 & 0.57 & -0.40 & -0.20 \\
        & $C_L$ (1/2 cycle) & 1.02 & 2.08 & -0.62 & -0.67 & 0.23 \\
        \bottomrule
    \end{tabular}
    }
    \caption{Decomposition of lateral force and thrust into components associated with difference in geometry of fin and that due to difference in pressure distribution.}
    \label{tab:example_table}
\end{table}

\section{Discussion}
Local redirection of pressure force due to fin deformation provides the solitary basis for the increased efficiency of the flexible fins and the alternative hypothetical mechanisms involving changes in \replaced{yaw}{pitch}ing moments associated with pressure distributions (Figs. \ref{fig:finComparison}(i)C, \ref{fig:finComparison}(ii)C) and changes in relative-phase between pressure-induced force and velocity (Figs. \ref{fig:phase}B--C, \ref{fig:flexibility_comparison}B) do not substantially contribute to improved efficiency. 

The efficiency gains provided by fin flexibility may be further understood by considering the power requirements for propulsion.
The power for each elemental area on the fin mesh (Fig. \ref{fig:SchemaModel}D) may be approximated by the product of its pressure difference ($\Delta p$) acting along the normal $\hat{n}$, area ($dS$) and velocity ($\vec{V}$).
The total fin power at a given time may therefore be found by integrating this product over the area of the fin, i.e. $P(t) \approx \int_{S_\text{fin}} \Delta p \, \hat{n} \cdot \vec{V} dS$.
This relationship underscores the crucial role of the timing of fin deformation (affecting the direction of $\hat{n}$) with respect to pressure and velocity changes.  During a tail-beat cycle, the fin experiences significant variations in the net pressure differential across its surface (Fig. \ref{fig:phase}B). The extremes in pressure occur shortly after the mid-point of the stroke, coinciding roughly with the highest fin velocity (Fig. \ref{fig:phase}C), which serves as a multiplicative factor in generating power (Eqn. \ref{eq:power-estimate-flexible}). 
% The peaks in fin deformation for the flexible fins occur close in time to the peaks in pressure differential.
The timing of fin deformation relative to the pressure differential (Fig. \ref{fig:flexibility_comparison}B,D) also further aids in improving the thrust performance of the flexible fins.
In particular, extreme deformations in the fore-aft orientation roughly coincide with peak pressure to enhance thrust (Fig. \ref{fig:phase}B, G).
Deformations in the dorso-ventral orientation proceed with similar timing (Fig. \ref{fig:phase}K) which serve to redirect lateral forces into directions that do not contribute to the power requirements (Eqn. \ref{eq:power-estimate-flexible}). In fact, the lateral orientation of the most flexible fin is nearly minimal at these phases, which serves to reduce the generation of energetically costly lateral forces (Fig. \ref{fig:phase}C, H).

Therefore, local-force redirection creates a matching of the timing of pressure, velocity and deformation changes in a manner that reduces power and enhances propulsive efficiency.
% These pressure differentials across the fin surface in turn, serve as the primary drivers of fin deformation and so, the phase alignment between the pressure differential and the dorso-ventral deformation is also not a coincidence. 
This seemingly complex combination of physical phenomena enables efficient locomotion primarily through passive interactions between the fin’s mechanical design and the surrounding flow. Similar levels of performance would otherwise require active sensing, neural integration, and finely coordinated muscular control. By bypassing these energetically costly and potentially less robust control mechanisms, and instead exploiting fluid–structure interactions to enhance performance, the flexible caudal fin emerges as a compelling example of mechanical intelligence.

We note that the caudal fin can experience antero-posterior (or chordwise) deformation and\/or dorso-ventral deformation. Several studies that employed 2D \cite{bergmann2014effect} or 3D models of BCF swimming \cite{seo2022improved, borazjani2009numerical, borazjanipof} have included chordwise bending\/deformation of the caudal fin in their models but have not included dorso-ventral deformation of the caudal fin. Thus, chordwise deformation is possible without dorso-ventral deformation. However, Bainbridge \cite{bainbridge1963caudal} identified and studied dorso-ventral deformation in caudal fins as a distinct phenomenon and our study is motivated by his observations. In our model, by excluding any chordwise deformation of the dorsal and ventral edges of the fin, we eliminate any \emph{extrinsic} chordwise deformation of the fin. The fin in our model therefore deforms in a dorso-ventral mode but of course, if we consider any horizontal section of the caudal fin, the dorso-ventral deformation also generates a chordwise deformation. However, this chordwise deformation would not exist in our model if we did not allow dorso-ventral deformation (as is the case for the stiff fin case) and we therefore, attribute any resulting consequences to dorso-ventral deformation and not to chordwise deformation.

These results also provide insight into Bainbridge’s conjecture regarding the benefits of fin flexibility \cite{bainbridge1963caudal}. Based on observations of the caudal fin’s ``billowing'' motion, Bainbridge hypothesized that dorso--ventral deformation in flexible fins smooths temporal variations in thrust, thereby promoting stable swimming. Our simulations show that, while the peak-to-peak variations in thrust are comparable across fins of differing flexibility (Fig.~\ref{fig:flexibility_comparison}B), oscillations in the larger \emph{lateral} force are substantially reduced for flexible fins (Fig.~\ref{fig:flexibility_comparison}C). This reduction in lateral force decreases the associated sway force and yaw moment induced by the fin, resulting in a smoother swimming motion. Collectively, these FSI simulations therefore support Bainbridge’s conjecture.

Propulsion with a flexible caudal fin bears similarities to the aerodynamics of wing deformations in flying animals. For example, 
these deformations improve the efficiency of flight in butterflies \citep{Zheng-Hedrick-Mittal2013} and bluebottle flies \citep{hsu2024wing}.
In addition, the flexibility of wings improves the lift generation during the clap-and-fling flight of butterflies \cite{johansson2021butterflies}. 
Thus, the current study offers an analogous example of how flexibility enhances propulsion in animal locomotion through the FSI dynamics.

The principle of local-force redirection for enhancing propulsive efficiency could be adopted in the design of efficient bioinspired propulsion systems. Such efficiency gains could enhance propulsion in autonomous underwater vehicles \cite{fish2003conceptual,triantafyllou2004review}, wave-assisted propulsion systems \citep{raut2024hydrodynamic, raut2025dynamics}, as well as flapping-foil based energy harvesting systems \citep{YOUNG20142,XIAO2014174}.

Lastly, while the prime focus of the work has been on the efficiency gains enabled by the caudal fin's flexibility, other fins, such as dorsal, anal, and pectoral, are also important in the swimming of fish. Tytell \cite{tytell2006median} experimentally analyzed the fin hydrodynamics and concluded that the combined force from anal and soft dorsal fin is comparable to that of a caudal fin. Thus the study could be expanded to model flexibility and FSI in these other fins as well with asymmetric deformation. However, this would increase the model complexity significantly and is out of scope for the current study. 

\section{Conclusions}
\label{sec:summary}
The computational modeling study presented here demonstrates that dorso-ventral deformations in flexible caudal fins significantly enhance the efficiency of propulsion by harnessing passive flow-induced deformation. Flexible fins deform dynamically under pressure loads, effectively reorienting hydrodynamic pressures to optimize thrust production while minimizing energetically costly lateral forces. Unlike rigid fins, which experience greater lateral forces and increased power demands, flexible fins exploit dorso-ventral deformation to passively and continuously adjust their geometry to fluid dynamic conditions, resulting in efficiency gains of up to 70\%. This form of mechanical intelligence highlights the potential for bioinspired designs that integrate compliant materials and structural flexibility to achieve improved energy efficiency and adaptability in engineered aquatic propulsion systems.

\ack
SK, RM, and JHS acknowledge support from ONR Grant N00014-22-1-2655. MJM acknowledges ONR Grant N00014-19-1-2035 and NSF grant IOS-2326484. The work benefited from the computational resources from the Advanced Research Computing at Hopkins (ARCH) facility at Johns Hopkins University. Thanks to Sean Ono for contributing the photo of the caudal fin of the batfish.

\bibliographystyle{unsrt}
\bibliography{main}

\appendix

\section{Grid Independence Study}\label{app:grid_indep}
\begin{figure}
    \centering
    \subfloat{\includegraphics[width=0.48\textwidth]{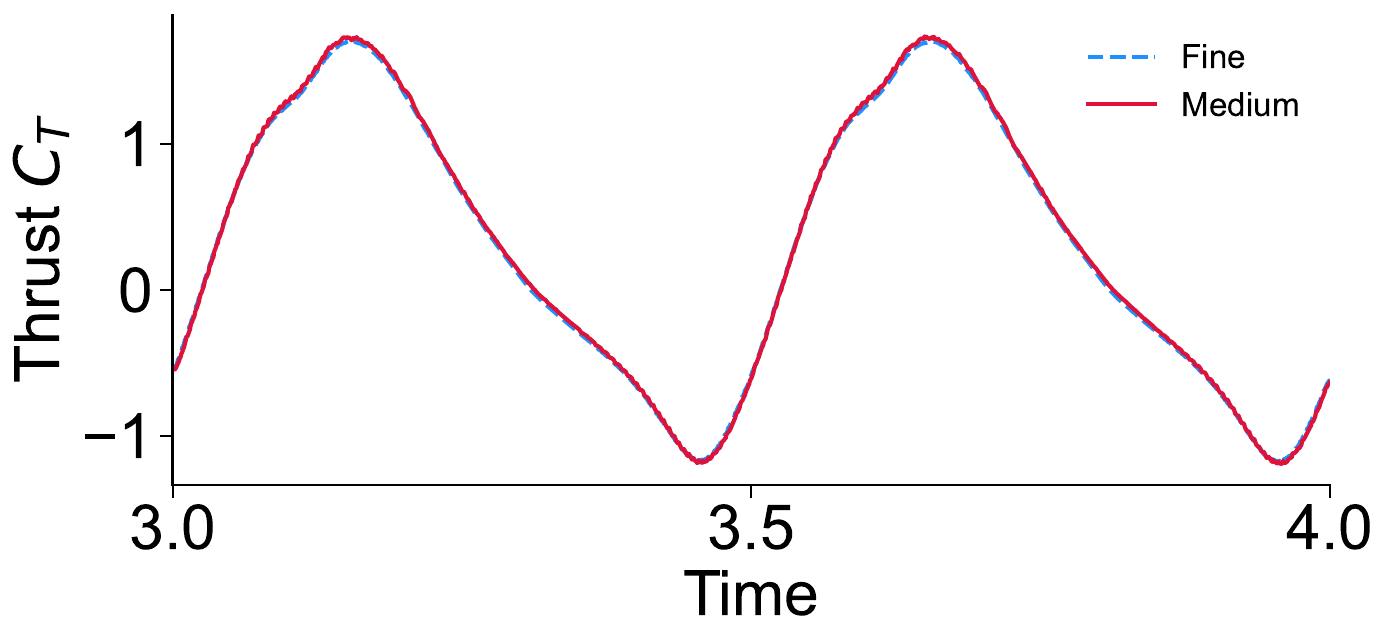}\label{fig:refinethrust}}
%    \hfill
    \subfloat{\includegraphics[width=0.48\textwidth]{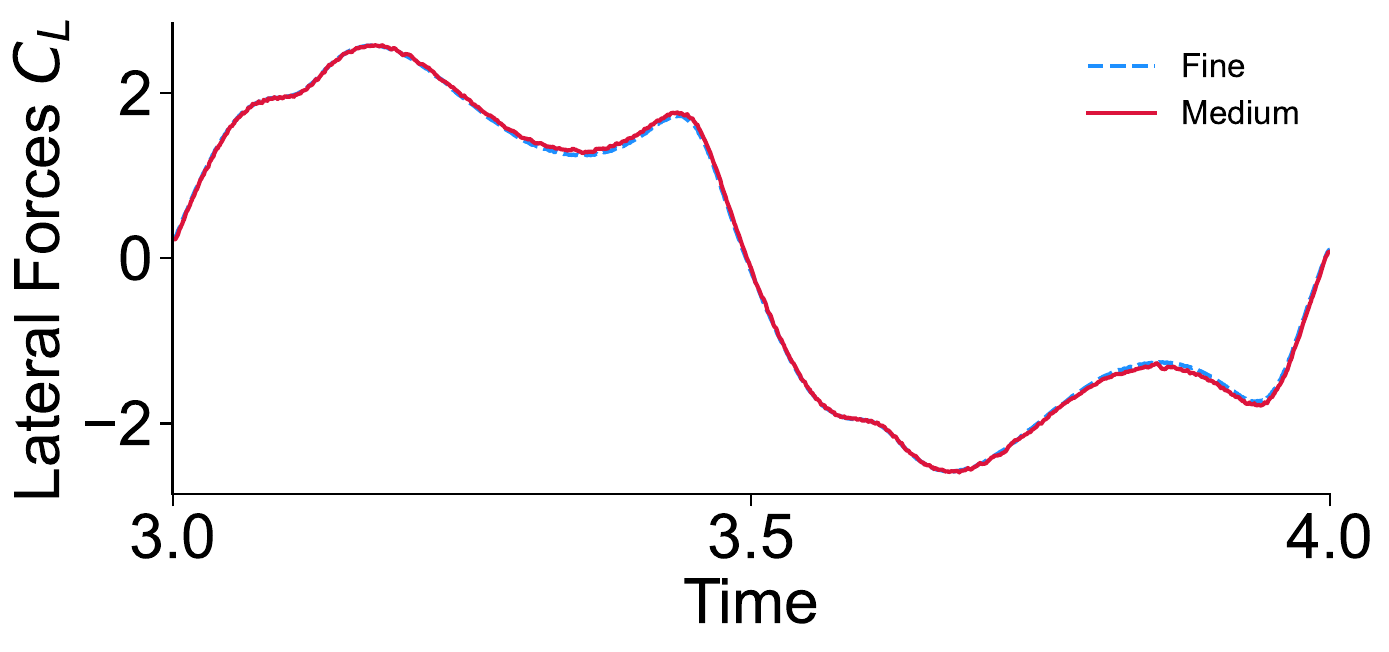}\label{fig:refinelift}}
    \caption{Grid refinement study results for a membrane with $h^* = 0.02$ at $St=0.3$. (a) Comparison of thrust force coefficient (b) Comparison of lateral force coefficient}
    \label{fig:gridRefinement}
\end{figure}

Here, we present our results for the grid independence study. We chose two grid resolutions termed ``Fine" and ``Medium" to show that the simulations performed are grid-independent. There are two regions in our rectilinear grid, one with constant grid space and another one with non-uniform grid spacing. The Medium grid is generated such that the minimum grid spacing $\Delta$ in the constant grid spacing regions $(\Delta x_1 = \Delta x_2 = \Delta x_3 = \Delta)$ is set to $0.01$ corresponding to 100 grid points along the fin chord. The stretching in the non-uniform grid is set to match the minimum grid spacing with a higher density of grid/low-stretching in the downstream region to resolve vortices. Next, the fine grid is generated with a minimum grid spacing of 0.005, corresponding to 200 grid points along the fin chord length. The time traces of thrust and lateral force coefficients are presented in figure \ref{fig:gridRefinement}, and qualitatively, the results are approximately similar. We compute the percent error between cycle-averaged thrust and half-cycle-averaged lift to quantify the error. We found the error in the thrust calculation to be about $3.5\%$ and in the lateral force calculation to be $1.1\%$. This shows a very minimal error between calculations from two grids. Therefore, we used the medium grid for all our simulations.

\section{Stationary State Study}
\begin{figure*}[t]
    \centering
    {\includegraphics[width=0.8\textwidth]{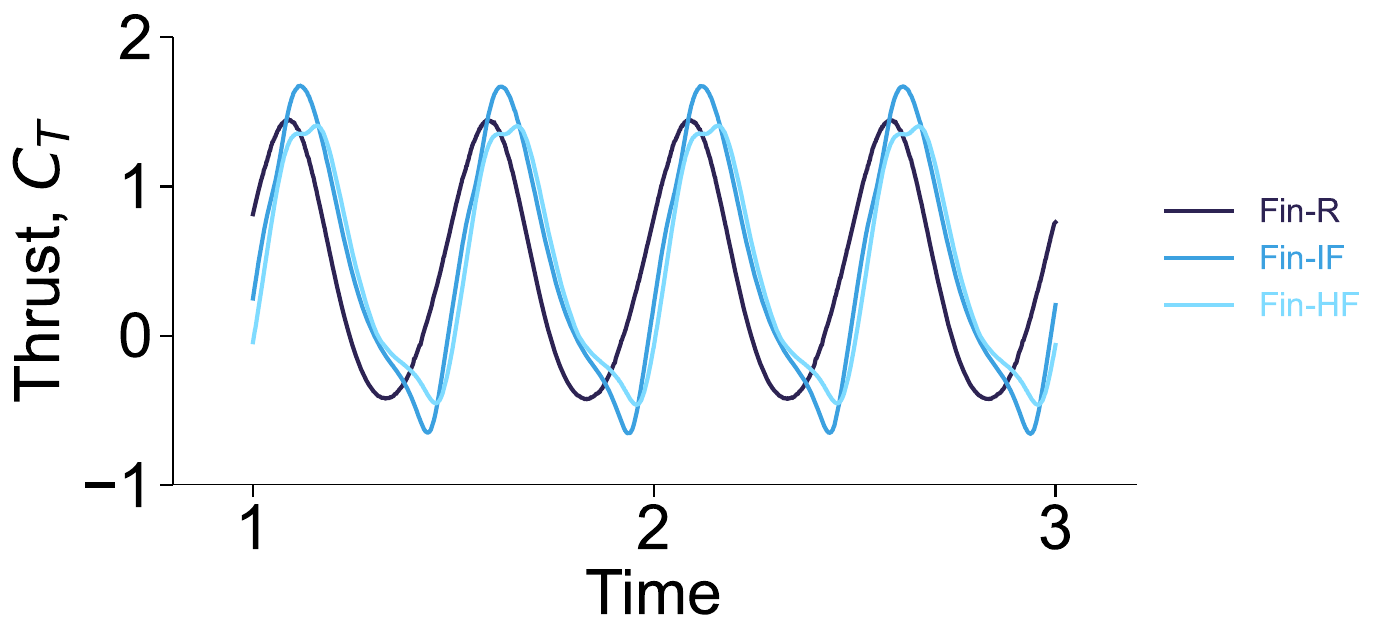}}
    \caption{Time-series data of thrust coefficient($C_T$) with normalized time ($t/T$) for second ($t/T = 1-2)$) and third cycle ($t/T = 2-3)$).
    }
    \label{fig:stationarystate} %time_series (i-k)
\end{figure*}

\added{
Here, we present our analysis to show that the simulations have reached a stationary state by the third flapping cycle. The stationary state is reached to ensure that the reported results are not affected by any transient effects in our simulations. We present this by showing the time-series data of thrust coefficient for the first two flapping cycles in figure \ref{fig:stationarystate}. The plot show that the time variation of thrust coefficient is identical for the two cycles. The error measured using the $L_2$ norm was found to be 0.5\%, 0.97\%, and 0.6\% for Fin-R, Fin-IF, and Fin-HF, respectively. These results show minimal differences between the second and third cycles, indicating that the flow has reached a stationary state.
}
\end{document}